\PassOptionsToPackage{unicode}{hyperref}
\PassOptionsToPackage{hyphens}{url}
\PassOptionsToPackage{dvipsnames,svgnames,x11names}{xcolor}
\documentclass[
]{article}
\usepackage{amsmath,amssymb}
\usepackage{lmodern}
\usepackage{iftex}
\ifPDFTeX
  \usepackage[T1]{fontenc}
  \usepackage[utf8]{inputenc}
  \usepackage{textcomp} % provide euro and other symbols
\else % if luatex or xetex
  \usepackage{unicode-math}
  \defaultfontfeatures{Scale=MatchLowercase}
  \defaultfontfeatures[\rmfamily]{Ligatures=TeX,Scale=1}
\fi
\IfFileExists{upquote.sty}{\usepackage{upquote}}{}
\IfFileExists{microtype.sty}{% use microtype if available
  \usepackage[]{microtype}
  \UseMicrotypeSet[protrusion]{basicmath} % disable protrusion for tt fonts
}{}
\makeatletter
\@ifundefined{KOMAClassName}{% if non-KOMA class
  \IfFileExists{parskip.sty}{%
    \usepackage{parskip}
  }{% else
    \setlength{\parindent}{0pt}
    \setlength{\parskip}{6pt plus 2pt minus 1pt}}
}{% if KOMA class
  \KOMAoptions{parskip=half}}
\makeatother
\usepackage{xcolor}
\usepackage{graphicx}
\makeatletter
\def\maxwidth{\ifdim\Gin@nat@width>\linewidth\linewidth\else\Gin@nat@width\fi}
\def\maxheight{\ifdim\Gin@nat@height>\textheight\textheight\else\Gin@nat@height\fi}
\makeatother
\setkeys{Gin}{width=\maxwidth,height=\maxheight,keepaspectratio}
\makeatletter
\def\fps@figure{htbp}
\makeatother
\providecommand{\tightlist}{%
  \setlength{\itemsep}{0pt}\setlength{\parskip}{0pt}}
\NewDocumentCommand\citeproctext{}{}
\NewDocumentCommand\citeproc{mm}{%
  \begingroup\def\citeproctext{#2}\cite{#1}\endgroup}
\makeatletter
 \let\@cite@ofmt\@firstofone
 \def\@biblabel#1{}
 \def\@cite#1#2{{#1\if@tempswa , #2\fi}}
\makeatother
\newlength{\cslhangindent}
\newlength{\csllabelwidth}
\newenvironment{CSLReferences}[2] % #1 hanging-indent, #2 entry-spacing
 {\begin{list}{}{%
  \setlength{\itemindent}{0pt}
  \setlength{\leftmargin}{0pt}
  \setlength{\parsep}{0pt}
  \ifodd #1
   \setlength{\leftmargin}{\cslhangindent}
   \setlength{\itemindent}{-1\cslhangindent}
  \fi
  \setlength{\itemsep}{#2\baselineskip}}}
 {\end{list}}
\usepackage{calc}

\ifLuaTeX
\usepackage[bidi=basic]{babel}
\else
\usepackage[bidi=default]{babel}
\fi
\babelprovide[main,import]{american}
\def\languageshorthands#1{}
\ifLuaTeX
  \usepackage{selnolig}  % disable illegal ligatures
\fi
\IfFileExists{bookmark.sty}{\usepackage{bookmark}}{\usepackage{hyperref}}
\IfFileExists{xurl.sty}{\usepackage{xurl}}{} % add URL line breaks if available
\hypersetup{
  pdftitle={dfcosmic: A Python package for cosmic ray removal},
  pdfauthor={Carter Lee Rhea, Pieter van Dokkum, Steven R. Janssens,
Imad Pasha, Roberto Abraham, William P. Bowman, Deborah Lokhorst, Seery
Chen},
  pdflang={en-US},
  colorlinks=true,
  linkcolor={Maroon},
  filecolor={Maroon},
  citecolor={Blue},
  urlcolor={Blue},
  pdfcreator={LaTeX via pandoc}}

\title{dfcosmic: A Python package for cosmic ray removal}

\definecolor{c53baa1}{RGB}{83,186,161}
\definecolor{c202826}{RGB}{32,40,38}

\usepackage[affil-it]{authblk}
\usepackage{orcidlink}
\author[1,2%
  ]{Carter Lee Rhea%
    \,\orcidlink{0000-0003-2001-1076}\,%
    }
\author[1,3%
  ]{Pieter van Dokkum%
    }
\author[1%
  ]{Steven R. Janssens%
    \,\orcidlink{0000-0003-0327-3322}\,%
    }
\author[1,3%
  ]{Imad Pasha%
    }
\author[1,4,5%
  ]{Roberto Abraham%
    }
\author[1,3%
  ]{William P. Bowman%
    \,\orcidlink{0000-0003-4381-5245}\,%
    }
\author[1,6%
  ]{Deborah Lokhorst%
    }
\author[1,4,5%
  ]{Seery Chen%
    }

\affil[1]{Dragonfly Focused Research Organization, 150 Washington
Avenue, Santa Fe, 87501, NM, USA%
  }
\affil[2]{Centre de Recherche en Astrophysique du Québec (CRAQ), Québec,
QC G1V 0A6, Canada%
  }
\affil[3]{Astronomy Department, Yale University, 219 Prospect St, New
Haven, CT 06511, USA%
  }
\affil[4]{David A. Dunlap Department of Astronomy \& Astrophysics,
University of Toronto, 50 St.~George Street, Toronto, ON M5S 3H4,
Canada%
  }
\affil[5]{Dunlap Institute for Astronomy \& Astrophysics, University of
Toronto, 50 St.~George Street, Toronto, ON M5S 3H4, Canada%
  }
\affil[6]{NRC Herzberg Astronomy \& Astrophysics Research Centre, 5071
West Saanich Road, Victoria, BC V9E 2E7, Canada%
  }
\date{01 February 2026}

\begin{document}
\maketitle

\section{Summary}\label{summary}

Astronomical images often show sharp features that are caused by cosmic
ray (CR) hits, hot pixels, or non-Gaussian noise. L.A.Cosmic
(\citeproc{ref-van_dokkum_cosmic-ray_2001}{van Dokkum, 2001}) is a
widely used edge detection algorithm that identifies and replaces such
features. Here we describe \texttt{dfcosmic}, a direct python port of
L.A.Cosmic utilizing PyTorch and C++ to enable efficient performance on
both CPUs and GPUs. The code was developed for the MOTHRA array, which
is projected to produce more than 1000 large format CMOS images every 15
minutes. Compared to previous python implementations, \texttt{dfcosmic}
achieves a speed gain of at least 20\%.

\section{Statement of need}\label{statement-of-need}

The Modular Optical Telephoto Hyperspectral Robotic Array (MOTHRA) uses
CMOS sensors rather than traditional CCDs. Modern CMOS detectors have
extraordinarily low noise but suffer from a relatively high number of
hot pixels and non-Gaussian noise (``salt-and-pepper'';
(\citeproc{ref-alarcon_scientific_2023}{Alarcon et al., 2023})).
Therefore, the data reduction pipeline for MOTHRA requires rapid bright
pixel identification and removal for tens of thousands of images every
night using only a single core (2 threads) per frame. Although several
implementations of L.A.Cosmic
(\citeproc{ref-van_dokkum_cosmic-ray_2001}{van Dokkum, 2001}) exist such
as \texttt{lacosmic}
(\citeproc{ref-bradley_larrybradleylacosmic_2025}{Bradley, 2025}) and
\texttt{astroscrappy}
(\citeproc{ref-robitaille_astropyastroscrappy_2025}{Robitaille et al.,
2025}), these implementations either deviate from the original algorithm
in order to achieve computational gains or do not run fast enough for
our usage. Importantly, experiments on the preliminary data taken by
MOTHRA have demonstrated that it is crucial to use the original
implementation (notably a true median filter rather than a separable
median filter) in order to capture all the CRs (or hot pixels or
salt-and-pepper non-Gaussian noise) without accidentally removing bright
stars. \texttt{dfcosmic} has already been adopted in the nightly
reduction pipeline for MOTHRA.

\section{State of the field}\label{state-of-the-field}

More broadly modern high frequency observatories are taking thousands of
images each night; Therefore, it is necessary to have a fast, and
reliable, implementation of the algorithm to reduce all the data in a
reasonable amount of time. Although several methods for detecting CRs in
astronomical images have been proposed (i.e. Zhang \& Bloom
(\citeproc{ref-zhang_deepcr_2020}{2020}), Pych
(\citeproc{ref-pych_fast_2003}{2003}), Xu et al.
(\citeproc{ref-xu_cosmic-conn_2023}{2023})), the most widely use
algorithm is the L.A.Cosmic algorithm
(\citeproc{ref-van_dokkum_cosmic-ray_2001}{van Dokkum, 2001}). There
currently exist other cosmic ray removal codes in Python based on the
L.A.Cosmic algorithm; notably \texttt{lacosmic}
(\citeproc{ref-bradley_larrybradleylacosmic_2025}{Bradley, 2025}) and
\texttt{astroscrappy}
(\citeproc{ref-robitaille_astropyastroscrappy_2025}{Robitaille et al.,
2025}). Moreover, although the current data reduction infrastructure
only supports CPU computing, we wish to have a package that will
eventually be able to run rapidly on a GPU.

In light of these considerations, we have developed \texttt{dfcosmic}.
In order to demonstrate the speed benefits of using \texttt{dfcosmic},
we benchmark it against the \texttt{astroscrappy} and \texttt{lacosmic}
implementations. Notably, \texttt{astroscrappy}'s default run
configuration uses a separable median filter which is a non-negligible
departure from the original algorithm in order to speed up the
algorithm. We run \texttt{astroscrappy} in our benchmarking with and
without the separable median filter active. We stress that the use of a
separable median filter can result in the incorrect removal of cosmic
rays by incorrectly classifying the center of near-saturated or
saturated stars as cosmic rays. We show the results of the three
different versions of \texttt{dfcosmic}: CPU with torch only, CPU with
C++ optimization, and GPU. We discuss these three different versions in
the software design section of this paper.

We run the codes on the mock data used for testing by
\href{https://github.com/astropy/astroscrappy/blob/main/astroscrappy/tests/fake_data.py}{\texttt{astroscrappy}}
with a typically sized frame for MOTHRA (4000x6500). Each option was run
employing 1, 2, 4, 8, and 16 threads. The GPU used in this test was an
NVIDIA GeForce RTX 5060 Ti 16GB while the CPU was an AMD Ryzen 9 9950X
16-Core Processor.

\begin{figure}
\centering
\includegraphics[keepaspectratio]{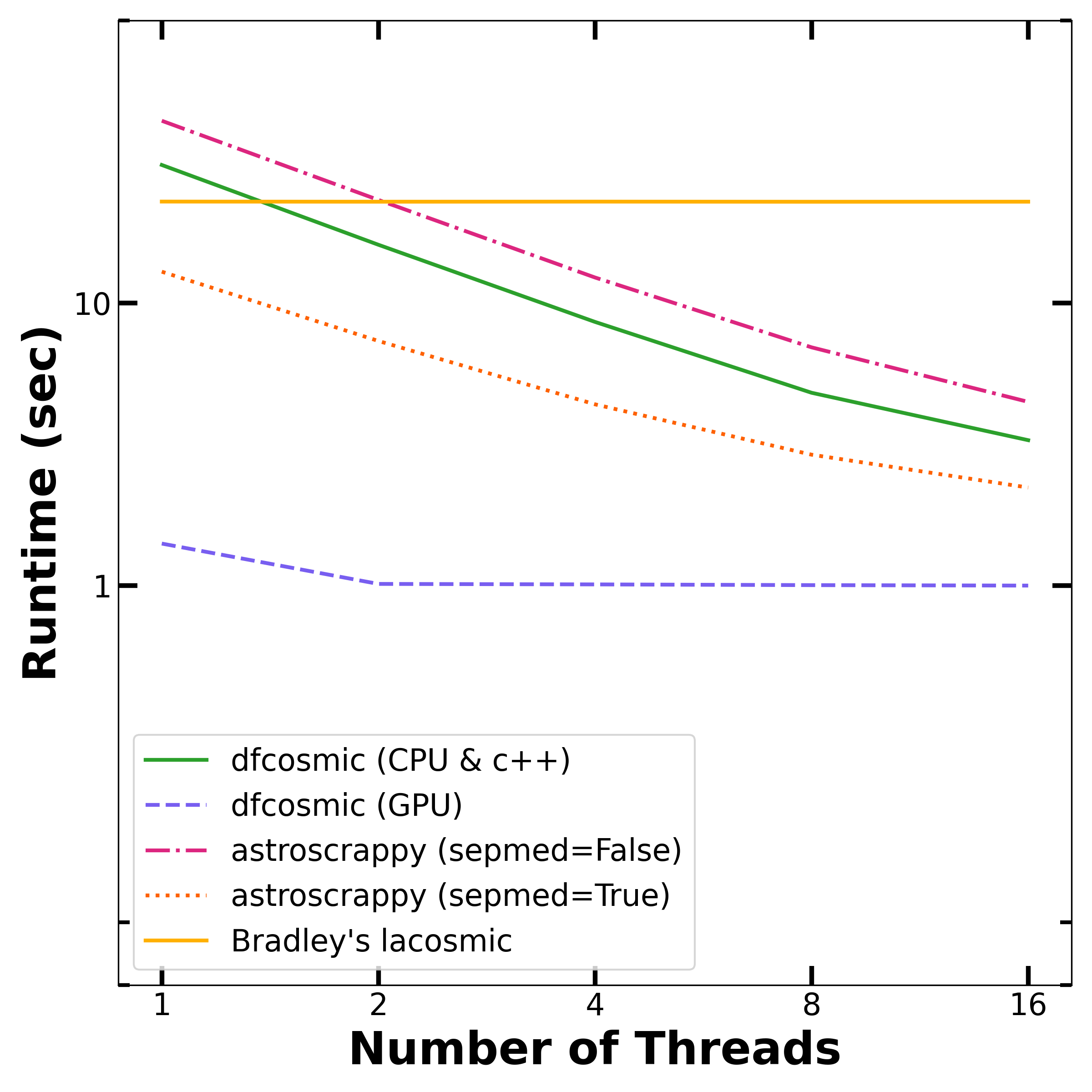}
\caption{\label{fig:comparison} Timing comparison of the many variants
of cosmic ray removal codes.}
\end{figure}

Although the performance gain with 2 or more threads may seem small
compared with the overall runtime, even a minimal gain for each
individual frame corresponds to a considerable gain when running the
pipeline on several tens of thousands of frames nightly. We reiterate
that \texttt{dfcosmic} always uses a true median
(i.e.~\texttt{sepmed=False}).

\section{Software design}\label{software-design}

\texttt{dfcosmic} was designed to be a simple PyTorch implementation of
the cosmic ray reduction algorithm initially developed in
(\citeproc{ref-van_dokkum_cosmic-ray_2001}{van Dokkum, 2001}). The code
was complexified in order to achieve greater reductions in speed.
Notably, initial benchmarking revealed the dilation and median filter
algorithms to be considerable bottlenecks. Therefore, we adopted C++
implementations of both of these algorithms. Doing so allowed
\texttt{dfcosmic} to be competitive (and even outperform) existing
implementations when run on 2 or more threads. We chose to use PyTorch
instead of a more standard library, such as numpy or scipy, so that we
could take advantage of the GPU, if available. As demonstrated in
\autoref{fig:comparison}, the GPU implementation is more than an order
of magnitude faster than any other implementation. Although not explored
here, the GPU implementation also allows for batch processing which can
enable further speedup.

In order to ensure the fidelity of the functions run internally, we
wrote custom torch implementations of the following:
\texttt{block\_replicate\_torch}, \texttt{convolve},
\texttt{median\_filter\_torch}, \texttt{dilation\_pytorch},
\texttt{sigma\_clip\_pytorch}. In case a user is unable to enable C++,
we allow the code to default to the custom torch implementations of the
median filter and dilation algorithms; we note that this leads to a
worse performance as compared to the C++ implementations.

\section{Research impact statement}\label{research-impact-statement}

\texttt{dfcosmic} is a new implementation of the well-known L.A.Cosmic
algorithm developed by (\citeproc{ref-van_dokkum_cosmic-ray_2001}{van
Dokkum, 2001}). \texttt{dfcosmic} is integrated into the nightly
reduction pipeline for the partially-constructed MOTHRA. The adoption of
this algorithm has considerable impact on the speed of reductions.
During a typical night, a single MOTHRA array takes approximately 1200
raw frames (this is the low end); note that the final version of MOTHRA
will have 30 arrays operating simultaneously. On the existing computing
infrastructure on AWS, the reduction of a frame using
\texttt{astroscrappy(sepmed=False)} using 2 threads per frame takes
approximately 75 seconds; using \texttt{dfcosmic} with the same setup
takes approximately 55 seconds. When run on all the frames taken in an
evening, this is equivalent to saving 24,200 seconds (6.7 hours) of
computing time for a single array.

\section{Methods}\label{methods}

\subsection{Algorithm}\label{algorithm}

The algorithm follows the methodology described in detail in
(\citeproc{ref-van_dokkum_cosmic-ray_2001}{van Dokkum, 2001}). Below, we
outline the main steps:

\begin{enumerate}
\def\labelenumi{\arabic{enumi}.}
\tightlist
\item
  Run laplacian detection
\item
  Create a noise model
\item
  Create significance map
\item
  Compute initial cosmic ray candidates
\item
  Reject compact, underampled objects (i.e.~stars or HII regions)
\item
  Determine which neighboring pixels to include
\item
  Replace cosmic rays with median of neighbors
\end{enumerate}

Importantly, we use the classic median filter rather than any optimized
version. We overcome the additional computational costs associated with
this computation by implementing our methodology in \texttt{PyTorch}
(\citeproc{ref-paszke_pytorch_2019}{Paszke et al., 2019}) with certain
functions (the median filter and dilation) written in C++.

\subsection{Main parameters}\label{main-parameters}

There are several key parameters that a user can set depending on their
specific use case:

\begin{enumerate}
\def\labelenumi{\arabic{enumi}.}
\tightlist
\item
  \texttt{objlim}: the contrast limit between cosmic rays and underlying
  objects
\item
  \texttt{sigfrac}: the fractional detection limit for neighboring
  pixels
\item
  \texttt{sigclip}: the detection limit for cosmic rays
\end{enumerate}

Furthermore, the user can supply the gain and readnoise. If a gain is
not supplied, then it will be estimated at each iteration.

\section{Results}\label{results}

\subsection{Example}\label{example}

In order to showcase \texttt{dfcosmic}, we apply it, along with
\texttt{astroscrappy} and \texttt{lacosmic} implementations, to the
original example from (\citeproc{ref-van_dokkum_cosmic-ray_2001}{van
Dokkum, 2001}) of the \emph{HST} WFPC2 image of galaxy cluster MS
1137+67.

\begin{figure}
\centering
\includegraphics[keepaspectratio]{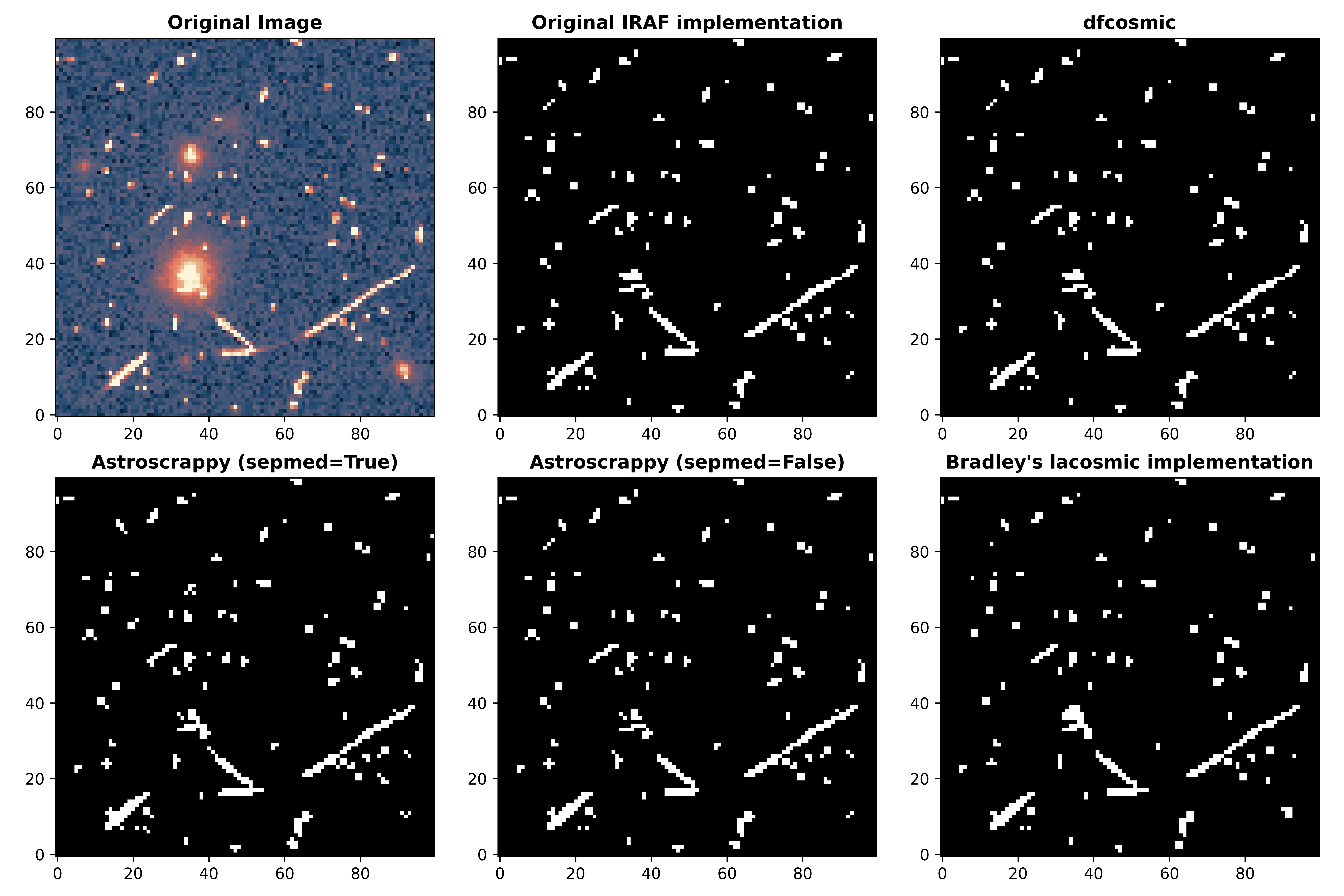}
\caption{\label{fig:demo} \emph{HST} WFPC2 image of galaxy cluster MS
1137+67. In the top panel, we show the original image, the mask from the
original IRAF implementation, and the mask from \texttt{dfcosmic}. In
the bottom row, we show the mask created by \texttt{astroscrappy} with
the \texttt{sepmed} argument set to True (left) and False (middle) while
in the right panel we show the mask from \texttt{lacosmic}
(\citeproc{ref-bradley_larrybradleylacosmic_2025}{Bradley, 2025}).}
\end{figure}

As demonstrated in \autoref{fig:demo}, the only Python implementation
that replicates the mask from the original IRAF implementation is
\texttt{dfcosmic}. By comparison, the other two popular implementation
either underestimate (\texttt{astroscrappy}) or overestimate (Bradley's
\texttt{lacosmic}) the size of the CRs in stars. An incorrect masking of
CRs in these regions can have a profound effect on the measured stellar
photometries.

\section{AI usage disclosure}\label{ai-usage-disclosure}

Generative AI was used for two aspects of this project:

\begin{enumerate}
\def\labelenumi{\arabic{enumi}.}
\item
  Claude.ai was used to help write/augment the unit tests and understand
  the original IRAF implementation.
\item
  ChatGPT was used to write the C++ code for the median filter function
\end{enumerate}

All code produced by AI was manually inspected for correctness.

The core functionality was not influenced by AI.

\section{Acknowledgements}\label{acknowledgements}

We acknowledge the Dragonfly FRO and particularly thank Lisa Sloan for
her project management skills.

We use the cmcrameri scientific color maps in our demos
(\citeproc{ref-crameri_scientific_2023}{Crameri, 2023}).

\section*{References}\label{references}
\addcontentsline{toc}{section}{References}

\protect\phantomsection\label{refs}

\end{document}